# Scaling Phononic Quantum Networks of Solid-State Spins with Closed Mechanical Subsystems

Mark C. Kuzyk and Hailin Wang

Department of Physics, University of Oregon, Eugene, OR 97403, USA

## Abstract

Phononic quantum networks feature distinct advantages over photonic networks for on-chip quantum communications, providing a promising platform for developing quantum computers with robust solid-state spin qubits. Large mechanical networks including one-dimensional chains of trapped ions, however, have inherent and well-known scaling problems. In addition, chiral phononic processes, which are necessary for conventional phononic quantum networks, are difficult to implement in a solid-state system. To overcome these seemingly unsolvable obstacles, we have developed a new network architecture that breaks a large mechanical network into small and closed mechanical subsystems. This architecture is implemented in a diamond phononic nanostructure featuring alternating phononic crystal waveguides with specially-designed bandgaps. The implementation also includes nanomechanical resonators coupled to color centers through phonon-assisted transitions as well as quantum state transfer protocols that can be robust against the thermal environment.

# I. INTRODUCTION

Photons are excellent carriers of quantum information and are the ideal choice for long distance quantum communications and networks[1-5]. For on-chip communications and networks, there are, however, a few inherent limitations. For example, the speed of light can be too fast for communications over short distances, such as a few hundred micrometers or less. Scattering losses of electromagnetic waves into vacuum can be excessive even with state-of-the-art nanofabrication technologies, which severely limits the photon lifetime in nano-optical systems such as photonic crystal optical resonators.

In comparison, phonons, which are the quanta of mechanical waves, feature several distinct advantages for on-chip communications[6-9]. The speed of sound is about five orders of magnitude slower than the speed of light. Mechanical waves cannot propagate in vacuum and are thus not subject to scattering or radiation losses into vacuum. The relatively long acoustic wavelength also makes it easier to fabricate phononic nanostructures for confining and guiding acoustic waves on a chip. Note that trapped ions, one of the most successful platforms for quantum computing, can be viewed effectively as a microscopic phononic network, with complete quantum control of both spin and motional degrees of freedom [10].

Nevertheless, there are two inherent obstacles in scaling up a phononic network. First of all, the coupling rate between a qubit and a mechanical mode at the single-phonon level scales with the zero-point fluctuation of the mechanical system, which is proportional to $1/\sqrt{m}$, with $m$ being the mass of the mechanical system. The larger the mechanical network is, the smaller the single-phonon coupling rate becomes. Secondly and more seriously, nearest neighbor coupling of a large number of mechanical resonators leads to the formation of spectrally-dense mechanical modes. The crosstalk between the collective mechanical modes prevents the quantum control of individual mechanical modes. These scaling problems are well known in ion trap quantum computers[11], for which phonon-mediated interactions play an essential role. In particular, the second scaling problem severely limits the number of ions that can be used in an ion trap module. Ion shuttling and photonic quantum channels between the modules have been pursued for scaling up ion trap quantum computers[3,11]. Various approaches have also been proposed to tackle the scaling problems of a solid-state system[9,12].

The primary function of a quantum network is to enable high-fidelity quantum state transfer between neighboring quantum nodes. This can take place in a cascaded network[13], for which the coupling between the neighboring nodes is unidirectional. Cascaded optical quantum networks can be realized with chiral optical interactions[14,15], as demonstrated with atoms and quantum dots. The lack of easily accessible chiral acoustic processes, however, makes it difficult to implement cascaded phononic quantum networks[6,9].

The scaling problems discussed above are unavoidable in a large mechanical network. To overcome these problems, we propose a general and conceptually-simple approach, which breaks the large mechanical network into small and closed mechanical systems. The use of these closed mechanical subsystems not only overcomes the scaling problems, but also avoids the technical difficulty of implementing chiral phononic processes. To realize this new approach, we use alternating waveguides and employ two waveguide modes for communications between neighboring quantum nodes. As illustrated schematically in Fig. 1a, each mechanical resonator couples to two distinct phononic waveguides, *A* and *B*, which allow phonon propagation at frequencies near $\omega_A$ and $\omega_B$, but forbid phonon propagation at frequencies near $\omega_B$ and $\omega_A$, respectively. This special feature of the waveguides can make any two neighboring resonators and the waveguide between them a closed mechanical subsystem, as highlighted in Fig. 1a. For a phononic quantum network of solid state spins, we use quantum nodes, in which a spin system couples selectively to two mechanical resonator modes with frequency $\omega_A$ and $\omega_B$, respectively. This phononic quantum network can be viewed as closed mechanical subsystems coupled together via the spins, as shown schematically in Fig. 1b. In this network, high-fidelity quantum state transfer between the neighboring spin systems takes place via the closed mechanical subsystems.

We describe an implementation of this architecture employing diamond color centers, nanomechanical resonators, and phononic crystal waveguides. In this implementation, color centers featuring robust spin qubits couple to vibrations of nanomechanical resonators through sideband (i.e. phonon-assisted) transitions driven by external optical or microwave fields[16]. Communications between these spin-mechanical resonators take place via alternating phononic crystal waveguides[17]. A key feature of the network is specially-designed phononic bandgaps in the phononic crystal waveguides, which enable the formation of closed mechanical subsystems. In addition, the entire network can be embedded in a phononic crystal lattice, which isolates and protects the network from the surrounding mechanical environment. Note that diamond photonic crystals and optomechanical crystals, which are technically more demanding than diamond phononic crystals in terms of nanofabrication, have already been successfully realized[18-20].

We also outline two schemes for quantum state transfer between spin systems in neighboring resonators. One scheme relies on spin-mechanical coupling of a single spin. The other employs spin ensembles for the quantum state transfer and approximates the spin ensemble as a bosonic oscillator[2,21,22]. Both schemes can be robust against the overall thermal environment. Fidelities exceeding 0.99 for quantum state transfer between single spins are achievable with current technologies.

Solid state spin systems such as negatively-charged nitrogen vacancy (NV) centers in diamond have emerged as a promising qubit system for quantum information processing[23-25]. High fidelity quantum control of individual spin qubits via microwave or optical transitions has been well established. Photonic networks of NV centers have been proposed[26-28]. Coherent spin-spin coupling mediated by electro-mechanical transducers or by dark spin chains has also been considered theoretically[12,29]. The phononic quantum network described in this paper can potentially enable a scalable, chip-based experimental platform for developing quantum computers using robust solid-state spin qubits.

## II. PHONONIC QUANTUM NETWORKS

The proposed phononic network consists of diamond spin-mechanical resonators that couple spin qubits in diamond to relevant mechanical modes, phononic crystal waveguides with suitable bandgaps and waveguide modes, and a two-dimensional (2D) phononic crystal lattice that protects the mechanical modes involved in the phononic network. For numerical calculations, we assume that the phononic network is fabricated from a diamond membrane with a thickness of 300 nm. In addition to NV centers, other color centers in diamond, such as silicon vacancy (SiV) or germanium vacancy (GeV) centers[30-33], can also be used in the phononic network. High quality NV, SiV, and GeV centers can be created in diamond through ion implantation, followed by elaborate thermal annealing and surface treatment[34,35].

### A. Spin-mechanical resonators

The elementary unit or node in our quantum network is a spin-mechanical resonator, in which spin qubits couple to mechanical resonator modes in a thin, rectangular diamond plate. Calculations of mechanical normal modes in the diamond plate are discussed in detail in the appendix. We are interested in mechanical compression modes that are symmetric with respect to the median plane of the plate (the so-called symmetric modes). Figure 2a shows, as an example, the displacement pattern of a fifth order compression mode.

Coherent interactions between electron spin states of a NV center and long-wavelength mechanical vibrations of the diamond lattice have been experimentally explored via either ground-state or excited-state strain coupling[16,36-42]. The orbital degrees of freedom of a NV center can couple strongly to the long-wavelength mechanical vibrations via the excited states. As a result, the excited-state strain coupling for a NV center is about five orders of magnitude stronger than the ground-state strain coupling[42-44]. For defect centers such as SiV and GeV centers, strong coupling between the orbital degrees of freedom and the mechanical vibrations can also take place through the ground states[45].

As illustrated in Fig. 2b, we control the coupling between the ground spin states of the NV center and the relevant mechanical mode through a resonant Raman process that consists of a sideband (or phonon-assisted) optical transition as well as a direct dipole optical transition. The Raman process is driven and controlled by two external optical fields. The interaction Hamiltonian is given by[40]

$$V = \hbar \frac{\Omega_-}{2} \frac{g_s}{\omega_m} (\hat{a} e^{i(\Delta_- - \omega_m)t} |e\rangle\langle -| + h.c.) + \hbar \frac{\Omega_+}{2} (e^{i\Delta_+ t} |e\rangle\langle +| + h.c.) , \qquad (1)$$

where $g_s = D k_m x_{zpf}$, $D$ is the deformation potential, $x_{zpf}$ is the zero-point fluctuation, $k_m$ is the phonon wavevector, $\Omega_+$ and $\Omega_-$ are the optical Rabi frequencies and $\Delta_+$ and $\Delta_-$ are the effective dipole detunings for the two respective optical transitions, and $\hat{a}$ is the annihilation operator for a mechanical mode with frequency $\omega_m$. For a NV center, the $m_s = \pm 1$ ground spin states can serve as states $|\pm\rangle$ and the $A_2$ state can serve as state $|e\rangle$ [46]. A detailed derivation of Eq. 1 is given in the appendix of Ref. [40].

The use of the sideband transitions, instead of resonant transitions, enables the selective coupling of an electron spin to any relevant mechanical modes, including two or more mechanical modes, which is an essential requirement for the implementation of the proposed network architecture. Specifically, we can couple the electron spin states to a mechanical mode with frequency $\omega_m$ by setting the detuning between the two optical driving fields according to the Raman resonant condition, $\Delta_- - \omega_m = \Delta_+$.

A limitation of using the resonant Raman process for coherent spin-mechanical coupling is the optically-induced decoherence due to the optical excitation of the excited state, with a decoherence rate given by $\gamma_{opt} = (\Omega/2\Delta)^2 \Gamma_{ex}$, where $\Omega$ is the optical Rabi frequency, $\Delta$ is the relevant dipole detuning, and $\Gamma_{ex}$ is the excited-state population decay rate. To suppress optically-induced decoherence, we can exploit a combination of techniques, such as dark states, shortcuts to adiabatic passage, and systematically-corrected control pulses [47-49], in addition to the use of a relatively large dipole detuning. Excited-state mediated spin-mechanical coupling via a dark state has already been demonstrated in an earlier experimental study[40].

For negatively charged SiV or GeV centers that feature strong ground-state strain coupling, the spin-mechanical coupling can be driven by microwave sideband transitions between the ground spin states. The resonant Raman process discussed above is no longer needed. The coupling schemes discussed in an earlier study can also be adopted[9].

## B. Alternating phononic crystal waveguides and closed mechanical subsystems

We use phononic crystal waveguides, which are one-dimensional (1D) phononic crystals consisting of a periodic array of holes in a beam (see Fig. 3a), to network together a series of spin-mechanical resonators. In a simple picture, mechanical vibrations in a resonator excite propagating mechanical waves in the adjacent phononic waveguides[17]. Conversely, mechanical waves in the phononic waveguide also excite vibrations in the adjacent mechanical resonators.

For the formation of closed mechanical subsystems, we employ alternating phononic crystal waveguides. The building block of our phononic network is shown in Fig. 3a, in which a spin-mechanical resonator couples to two phononic waveguides, *A* and *B*, that feature an array of elliptical holes with different periods. As shown in Fig. 3b, the phononic band structure of each waveguide shows a sizable bandgap for the symmetric mechanical modes (which have spatial displacement patterns that are symmetric with respect to the median plane of the 2D structure). The center of the bandgap for waveguide *B*, which features a shorter period, is higher in frequency than that for waveguide *A*, which features a longer period. The two bandgaps have two non-overlapping spectral regions, as highlighted by the grey shaded regions in Fig. 3b. Waveguide modes in the upper and lower shaded regions can only propagate in waveguide *A* and *B*, respectively. We use waveguide modes and resonator modes with frequencies in these non-overlapping regions for quantum state transfer between spin systems in neighboring quantum nodes.

For the design shown in Fig. 3a, we choose two specific resonator modes. The fifth order compression mode with frequency $\omega_A/2\pi=1.6332$ GHz, which is in the bandgap of waveguide *B* (see Fig. 3b), couples resonantly to a mode in waveguide *A*. The third order compression mode with frequency $\omega_B/2\pi=0.9133$ GHz, which in the bandgap of waveguide *A* (see Fig. 3b), couples resonantly to a mode in waveguide B. In this case, any two neighboring resonators in the network and the waveguide between them can form a closed subsystem, effectively realizing the network architecture shown in Fig. 1.

For the closed mechanical subsystem, the relevant waveguide modes are discrete standing wave modes. For a relatively short waveguide, the frequency spacing of these modes can be large compared with other relevant frequency scales and the waveguide can thus behave like a single-mode mechanical oscillator. In this limit, we can treat the closed mechanical subsystem as a three-mode system (see the appendix for the effective Hamiltonian and for more detailed discussions). Numerical calculations of the normal modes of the closed subsystem and, in particular, the coupling rate, $g$, between the resonator and the waveguide modes are presented in the appendix. Depending on the specific design of the waveguides and resonators, $g/2\pi$ can range from 1 MH to more than 10 MHz. We can also engineer the coupling rate by tailoring or shaping the contact area between the waveguide and the resonator.

It should be added that in the limit of a single-mode waveguide, the phononic network is effectively a network of mechanical resonators. In this limit, the phononic crystal structures are not necessary for the formation of closed mechanical subsystems. In practice, the relevant phononic bandgaps ensure the nearly complete suppression of the unwanted mechanical coupling between the neighboring resonators. We also note that the resonant conditions discussed above are required only for the adjacent resonator and waveguide modes, but not for distant resonator and waveguide modes, which implies less stringent requirements on the nanofabrication process.

### C. Isolating intra-node spin-mechanical coupling from the waveguides

We separate the spin qubits in a spin-mechanical resonator into computation qubits and communication qubits that are used exclusively for quantum state transfer between neighboring quantum nodes. These qubits can be addressed separately due to their different physical locations or their different frequencies or polarization selection rules. The computation qubits can greatly enhance the capability and especially the fault tolerance of the quantum network. For example, it has been shown recently that with 15 qubits of all-to-all connectivity in each quantum node, fault tolerant threshold can be reduced to 0.12%[50]. For our architecture, the all-to-all connectivity can be achieved via the intra-node interactions.

Ideally, intra-node interactions should be decoupled from the phononic waveguides, which can be accomplished by exploiting the bandgaps of the phononic crystal waveguides. Specifically, the computation qubits can couple to each other and to the communication qubits through a resonator mode with a frequency that is in the bandgap of both phononic crystal waveguides, i.e. in the overlapping spectral region of the two phononic bandgaps, as highlighted by the yellow shaded area in Fig. 3b. In this case, the phonon-mediated coupling among the computation qubits and the communication qubits within a spin-mechanical resonator is decoupled from the adjacent waveguides. For the resonator-waveguide design shown in Fig. 3a, the fourth order compressional mode of the resonator, with $\omega_C$=1.3258 GHz, falls in the bandgap of both phononic crystal waveguides and can thus serve as a mechanical mode for intra-node spin-mechanical coupling. Other resonator modes in the overlapping region of the two bandgaps can also be used for this purpose, providing flexibility in the physical location of the computation qubits.

**D. Protecting phononic networks with a 2D phononic crystal lattice**

To protect the relevant mechanical modes from the surrounding mechanical environment, we embed the entire phononic network in a 2D phononic crystal lattice, as illustrated in Fig. 4. 2D phononic crystal lattices have been used extensively in earlier studies to isolate mechanical systems such as optomechanical crystals, membranes, and single-mode phononic wires from the surrounding environment [51-54]. The use of 2D phononic crystal shields has led to the experimental realization of ultrahigh mechanical Q-factors, with $\omega Q_m/2\pi$ approaching or even exceeding $10^{17}$ [53,54].

The bottom panel of Fig. 4 plots the phononic band structure of the symmetric mechanical modes (i.e. modes that are symmetric about the median plane of the 2D structure) in the 2D phononic crystal lattice shown in the top panel. The band structure features a bandgap between 0.85 and 2.25 GHz, spanning the phononic bandgaps of both phononic crystal waveguides *A* and *B* and thus protecting all the mechanical modes relevant to the phononic quantum network. For the design shown in Fig. 4, only waveguide *B* is attached to the 2D lattice, because this waveguide and the 2D square lattice have the same period. In this case, mechanical modes with frequencies near $\omega_A$ are isolated from the environment by the bandgap in the 2D lattice as well as the bandgap in waveguide B, which also relaxes the requirement that the bandgap of the 2D lattice spans both $\omega_A$ and $\omega_B$.

The specific design for the mechanical resonators, phononic crystal waveguides, and 2D phononic crystal shields discussed above is by no means optimal. The design serves as an example for implementing the proposed network architecture in a phononic network.

### III. QUANTUM STATE TRANSFERS

Mechanically-mediated quantum state transfers have been investigated theoretically for optomechanical transducers that can interface hybrid quantum systems[55-60], as well as for pure mechanical systems[8,9,21,22]. State transfer processes that can be robust against thermal mechanical noise have also been proposed. One approach is based on the use of dark modes, which are decoupled from the relevant mechanical system through destructive interference[61,62]. Dark modes in multimode optomechanical and electromechanical systems have been realized experimentally [63-66]. Another approach returns the mediating mechanical mode to its initial state, disentangling the mechanical mode from the rest of the system[67-70].

The closed mechanical subsystem discussed above consists of three mechanical modes, including two resonator modes in the respective mechanical resonators, described by annihilation operators, $\hat{a}_1$ and $\hat{a}_2$, and a waveguide mode, described by $\hat{b}$. For simplicity, we assume that the two resonator modes couple to the waveguide mode with equal coupling rate *g* and all three mechanical modes have the same resonance frequency, unless otherwise specified. Each resonator mode couples to either a single spin or an ensemble of spins. For the quantum state transfer between the two spin systems in the respective resonators, the interaction Hamiltonian is given by,

$$H_I = \hbar g \hat{b}^+ (\hat{a}_1 + \hat{a}_2) + \hbar[G_1(t)\hat{S}_1 \hat{a}_1^+ + G_2(t)\hat{S}_2 \hat{a}_2^+] + h.c., \qquad (2)$$

where $\hat{S}_1$ and $\hat{S}_2$ describe the spin systems, as will be discussed in more detail later, and $G_1(t)$ and $G_2(t)$ are the corresponding spin-mechanical coupling rates. Spin qubits in a resonator can selectively couple to given mechanical modes of the resonator. As discussed earlier, the mode selection for the spin-mechanical coupling is set by the detuning between the external laser driving fields or by the frequency of the microwave driving field.

The relevant mechanical modes in the two resonators can be cooled to near their motional ground states. This can be achieved via resolved sideband cooling using a phonon-assisted optical transition[43], along with cryogenic cooling. Because of the protection provided by the 2D phononic crystal shield, the mechanical damping rate can in principle be much smaller than the relevant coupling rate such that mechanical losses can be ignored during the transfer process. With $(G_1, G_2) \gg k_B T / \hbar Q_m$, the effects of thermal heating during the transfer process can also be negligible. For T=1 K and $G_1$ and $G_2$ on the order of 0.1 MHz, this requires $Q_m \gg 10^5$, a regime readily achievable in state-of-the-art phononic nanostructures[53,54].

We consider two quantum state transfer schemes based on the use of single spins and spin ensembles, respectively. Both schemes can return the relevant mechanical system to its initial state and can thus be robust against the mechanical thermal environment[22,67,69]. It might be possible to combine and take advantage of both single spins and spin ensembles (in separate resonators) for quantum state transfers. Different color centers can also be used in the same photonic network. In addition, our network architecture allows a high degree of parallelism. Relevant gate or state transfer operations can take place in parallel, when they are implemented simultaneously in respective mechanical subsystems.

**A. Quantum state transfer between single spins**

For the single-spin based transfer scheme, the spin operator in Eq. (2) corresponds to the lowering operator for a single spin, with $\hat{S} = \hat{\sigma} = |-\rangle\langle+|$. The single spin, which serves as a communication qubit, can be positioned near the node of the resonator mode (see Fig. 2a), where the spin-mechanical coupling reaches its maximum value. For the resonant Raman process shown in Fig. 2b, the effective spin-mechanical coupling rate for a single spin is given by $G = g_s \Omega_+ \Omega_- / (4|\Delta_+| \omega_m)$ [40]. With estimated D=5 eV [42,43] and $x_{zpf}$=0.75x10$^{-15}$ m, we have $G/2\pi$ =0.1 MHz, where we take $\Omega_+/2\pi = \Omega_-/2\pi$=0.6 GHz, $\Delta_+/2\pi$ =3 GHz, and $\omega_m/2\pi$=1 GHz. In this case, the magnitude of G is limited by the relatively large dipole detuning for the resonant Raman process. Much greater G can be achieved with SiV and GeV centers[45], for which ground-state strain coupling can be used.

As shown in Fig. 5, the state transfer between the two spin systems can take place in a simple triple-swap process. For the first swap, we set $G_2=0$ and turn on $G_1$ for a duration $\tau_1=\pi/2G_1$, mapping the spin state for $\hat{S}_1$ to the state for $\hat{a}_1$. For the second swap, we set $G_1=G_2=0$. After a duration $\tau_2 = \pi/\sqrt{2}g$, the state of $\hat{a}_1$ is effectively mapped to that of $\hat{a}_2$ [68]. This waveguide-mediated mapping between the two mechanical resonators leaves the state in the waveguide unchanged, as shown in Fig. 5. For the third swap, we set $G_1=0$ and turn on $G_2$ for a duration $\tau_3=\pi/2G_2$, mapping the state from $\hat{a}_2$ to $\hat{S}_2$. Overall, the triple-swap process leads to an effective state swap between $\hat{S}_1$ and $\hat{S}_2$.

The unavoidable coupling to the waveguide mode during the swaps between the single spin and the resonator modes (i.e. the first and the third swap) limits the fidelity of the overall quantum state transfer, which is defined as

$$\mathcal{F} = \left( \mathrm{Tr}\left[ \left( \sqrt{\rho(t_f)} U_{\mathrm{SWAP}} \rho(t_i) U_{\mathrm{SWAP}}^+ \sqrt{\rho(t_f)} \right)^{1/2} \right] \right)^2, \qquad (3)$$

where $U_{\mathrm{SWAP}}$ represents the ideal two-qubit swap operation and $\rho$ is the reduced two-qubit density matrix [71]. Figure 6a shows the lower bound on fidelity, calculated over all initial spin states (using a dense mesh over the Bloch sphere) and with the initial mechanical state $|0,0,0\rangle$, as a function of $G/g$, where $G$ is the peak value for both $G_1$ and $G_2$. For the triple-swap process, high fidelity can be achieved only when $G/g \gg 1$. Figure 6b plots the fidelity when the duration of the $\pi/2$ pulses deviates from the ideal value. For relatively small $G/g$, the maximum fidelity actually occurs away from the zero deviation, $\varepsilon=0$. This is because for the phononic network, $g$ is a constant. The mechanical resonators remain coupled to the waveguide in the first and the third swap of the state transfer process. In the limit that $G/g \gg 1$, the maximum fidelity occurs at $\varepsilon=0$, as shown in Fig. 6b.

Detuning between the individual mechanical modes can also limit the fidelity of the state transfer. Here we assume that the single spin couples resonantly to the respective resonator mode since the corresponding detuning is set by the frequency of the driving lasers. Figure 6c shows the fidelity as a function of the detuning, $\delta_1$ and $\delta_2$, between the waveguide and the two resonators. As expected, high fidelity is achieved when the detuning is small compared with $g$. Figure 6d plots the fidelity with equal detuning for the two resonators, $\delta=\delta_1=\delta_2$.

In the limit that $G < g$, the spins in adjacent resonators can be coupled by any of the three normal modes of the resonator-waveguide-resonator system. In this case, the Molmer-Sorensen gate, which was initially developed for the trapped ion system[67], can be employed to achieve thermally-robust high-fidelity state transfer. Figure 7a shows schematically the coupling scheme of the Molmer-Sorensen gate between two single spins, for which each spin is driven bi-chromatically with opposite detuning $\pm\Delta_{MS}$ and with $G_1=G_2=G$. For $G=|\omega_m-\Delta_{MS}|/(2\sqrt{2K})$ and for a duration given by $2\pi K/|\omega_m-\Delta_{MS}|$, where $K$ is a positive integer, the Molmer-Sorensen gate drives the two spins from state $|--\rangle$ to a maximally entangled state $(|--\rangle -i|++\rangle)/\sqrt{2}$ (assuming only one mechanical normal mode is involved), which is also disentangled from the relevant mechanical mode[72]. State transfer between the two spins can be realized by combining the Molmer-Sorensen gate, which is a universal two-qubit gate, with additional single-qubit operations. Figure 7b plots the fidelity obtained at T=0.5 K as a function of $G$, with the mechanical modes initially in a thermal state and with $Q_m=10^7$, $\omega_m/2\pi=1$ GHz, and $\delta=0$. As discussed earlier, the high $Q_m$ should be achievable with the use of 2D phononic crystal shields and with the state-of-the-art nanofabrication technologies.

The Lindblad master equation used for the fidelity calculation includes noise sources from both the mechanical and spin systems. For the mechanical system, the noise includes the mechanical damping as well as the thermal noise. For the spin system, both the pure dephasing and population decay (or spin-flip) are included. The population decay rate is typically orders of magnitude smaller than the spin dephasing rate. For convenience, we take population decay time, $T_1=1$ $s$. As shown in Fig. 7b, a fidelity greater than 0.99 can be achieved with $G/2\pi$ exceeding 0.32 and 0.23 MHz for spin dephasing time $T_2^*=150$ and 250 $\mu s$ (for isotopically purified diamond[73]), respectively. The relatively large spin-mechanical coupling rate can be achieved with ground-state strain coupling of SiV or GeV centers[74].

For phononic quantum networks of single spins, the Molmer-Sorensen protocol, which operates in the regime of $G \ll g$, is preferable over the triple-swap protocol that requires $G \gg g$ for high-fidelity operations. The regime of $G \gg g$ will be difficult to achieve in diamond without compromising resonator-waveguide coupling. Furthermore, the Molmer-Sorensen protocol should also tolerate small detunings between the waveguide and resonator modes.

## B. Quantum state transfer between spin ensembles

For the spin-ensemble based transfer scheme, the spin operator in Eq. (2) corresponds to the collective lowering operator for a spin ensemble, with

$$\hat{S} = \frac{1}{\sqrt{<\sum_m (|-><-|-|+><+|)_m>}} \sum_m \hat{\sigma}_m, \qquad (4)$$

where we assume that the spins are initially prepared in the $|->$ state through optical pumping and the expectation value is taken with respect to the fully polarized state $\otimes_m |->$. Ground-state spin-strain coupling of SiV or GeV centers can be used to avoid large optical inhomogeneous broadening of the NV centers. Alternatively, a relatively large optical dipole detuning, $\Delta$, can be used for the ensemble NV centers. For sufficiently weak excitations, we can approximate $\hat{S}$ as a bosonic operator, with $[\hat{S}, \hat{S}^+] = 1$. Similar approximations for spin ensembles have also been used for thermally-robust quantum state transfer in an optical network[21,22]. In this limit, the overall system can be approximated as a set of linearly coupled harmonic oscillators.

With $G_1 = G_2 = G$, the interaction Hamiltonian can be written in terms of super modes, with $\hat{a}_\pm = (\hat{a}_1 \pm \hat{a}_2)/\sqrt{2}$ and $\hat{S}_\pm = (\hat{S}_1 \pm \hat{S}_2)/\sqrt{2}$, and with the form

$$H_I = \sqrt{2}\hbar g \hat{b}^+ \hat{a}_+ + \hbar G(\hat{S}_+^+ \hat{a}_+ + \hat{S}_-^+ \hat{a}_-) + h.c.. \qquad (5)$$

The corresponding Heisenberg equations can be solved analytically. The time evolution of $\hat{S}_1$ is given by

$$\hat{S}_1(t) = \frac{gG}{\Gamma^2}[\cos(\Gamma t) - 1]\hat{b} - \frac{iG}{\sqrt{2}\Gamma}\sin(\Gamma t)\hat{a}_+ - \frac{i}{\sqrt{2}}\sin(Gt)\hat{a}_-$$
$$+ \frac{1}{\sqrt{2}}[1 + \frac{G^2}{\Gamma^2}(\cos(\Gamma t) - 1)]\hat{S}_+ + \frac{1}{\sqrt{2}}\cos(Gt)\hat{S}_- \qquad (6)$$

where $\Gamma = \sqrt{2g^2 + G^2}$.

Under the condition that $\Gamma = 2nG$, where $n$ is a positive integer, $\hat{S}_1(t = \pi/G) = \hat{S}_2$, as can be seen from Eq. (6), and $\hat{S}_2(t = \pi/G) = \hat{S}_1$, which enables a perfect state transfer between the two spin systems, provided that $\hat{S}$ can be approximated as a bosonic operator. This state transfer process is independent of the initial states of the two mechanical resonators as well as the initial state of the phononic crystal waveguide.

To gain further physical insights into the quantum state transfer process, we plot in Fig. 8 the dynamics of the constituent mechanical and spin-ensemble systems under a constant spin-mechanical coupling. For simplicity, we assume that at $t=0$, the occupation in $\hat{S}_1$, $\hat{a}_1$, and $\hat{b}$ is 1 and that in $\hat{S}_2$ and $\hat{a}_2$ is 0. As shown in Fig. 8a (with $\Gamma = 2G$) and Fig. 8b (with $\Gamma = 4G$), an effective π-pulse (with duration $\tau = \pi/G$) swaps the quantum states of the two spin systems as well as those of the two mechanical resonator modes and returns the waveguide mode to its initial state. Because of the bosonic approximation of the spin ensembles, the dynamics of the constituent mechanical and spin systems are periodic. With $\Gamma = 2nG$, the complete state swapping between the two spin ensembles occurs simultaneously with that between the two mechanical resonator modes. This state swapping process, which arises from the periodic dynamics of the system, is independent of the phonon occupation or distribution in the individual mechanical modes (waveguide or resonator modes) and keeps the mechanical and the spin systems disentangled. In this regard, the state transfer can be robust against the overall thermal environment, provided that the thermalization rate, $k_B T / \hbar Q_m$, is small compared with other relevant coupling rates.

The above state transfer scheme requires a careful tuning of the spin-mechanical coupling rate, $G$, to satisfy the condition, $\Gamma = 2nG$. Nevertheless, the quantum state transfer process can tolerate considerable deviations of $G$ from its targeted or optimal value. As shown in Fig. 9a, even with a deviation as large as 3%, the lower bound of the fidelity for the state transfer process calculated with the effective Hamiltonian given in Eq. (5) can still exceed 0.99 (see the shaded area in Fig. 9a). Note that the lower bound in fidelity is computed in the same manner as Fig. 6.

In the limit that $\Gamma \gg G$ (which implies $G \ll g$), the fast dynamics of the "+" super-modes interacting with mode $\hat{b}$ effectively average to zero. As a result, the time evolution of mode $\hat{b}$ have negligible effects on the dynamics of the spin system, as shown in Fig. 8c. In this case, the time evolution can be described by the effective Hamiltonian

$$H_{eff} = \hbar G(\hat{S}_-^+ \hat{a}_- + \hat{S}_- \hat{a}_-^+). \tag{7}$$

The complete state swap between the two spin systems can now occur to the zeroth order of the small parameter $G/g$, with $\hat{S}_1(t = \pi/G) = \hat{S}_2$ and without the requirement that $\Gamma = 2nG$.

For the ensemble spin system, spin dephasing induced by the nuclear spin bath (which includes effects of inhomogeneous ground-state broadening) is a major limiting factor for the quantum state transfer process, especially in the regime of $G \ll g$. Figure 9b shows the fidelity for the state transfer as a function of the spin dephasing rate, $1/T_2^*$, calculated with the effective Hamiltonian given in Eq. (6) and with the mechanical parameters from the design presented in Appendix. The Lindblad master equation used includes effects of pure dephasing and population decay (with $T_1$=1 $s$). As expected, high fidelity can only be achieved when $1/T_2^*$ is small compared with $G$. In addition to the use of isotopically purified diamond[73], we can effectively suppress spin dephasing by using dressed, instead of, bare spin states[75,76].

We have also calculated the fidelity for the quantum state transfer of spin ensembles in a thermal environment. Similar to Fig. 7, the Lindblad master equation used includes noise sources from both mechanical and spin systems. The mechanical parameters used, $g/2\pi$= 5 MHz, $Q_m$=$10^7$, $\omega_m/2\pi$=1 GHz, and $\delta$=0, are derived from designs similar to those discussed in Appendix. With $G/2\pi$ =0.5 MHz and $T_2^*$=80 μs, we obtained a lower bound of 0.94 for fidelity obtained at a temperature of 100 mK. The fidelity increases to 0.98 at T=30 mK. A fidelity of 0.99 can be achieved if $T_2^*$ is increased to 160 μs (with T=30 mK). Note that this scheme is thermally less robust than the Molmer-Sorensen scheme discussed earlier. For the ensemble spin system, the key experimental challenge is to increase the number of spins, while minimizing the increase in the effective spin dephasing rate, because of possible interactions between the spins. We also note that effects of resonator-waveguide detuning, while not significant at relatively small fidelities, become noticeable when fidelities approach 0.99.

## IV. CONCLUSIONS

In summary, we have developed theoretically a phononic quantum network of solid state spins, in which a spin-mechanical resonator is coupled to two distinct phononic crystal waveguides. The specially-designed bandgaps in the alternating waveguides enable a new architecture for quantum networks. In this architecture, any two neighboring nodes and the waveguide between them can form a closed subsystem. This conceptually-simple architecture overcomes the inherent obstacles in scaling up phononic quantum networks and avoids the technical difficulty of employing chiral spin-phonon interactions. The proposed phononic quantum network thus provides a promising route for developing quantum computers that can take advantage of robust spin qubits.

We have considered two schemes for quantum state transfer between spin systems in neighboring quantum nodes, using single spins and spin ensembles, respectively. Both schemes can be robust against the thermal environment. These schemes are intended to illustrate examples of spin-mechanical interactions that can be used for the proposed phononic quantum networks. By using closed subsystems as building blocks, the phononic network can exploit and adopt a variety of quantum state transfer or entanglement schemes.

While the discussions in this paper use, as a specific example, color centers in diamond, the implementation can be applied or extended to other defect centers or solid-state spin systems such as SiC-based systems[77]. The general architecture in breaking a large network into small and closed subsystems and the specific approach of alternating, frequency-selective coupling can also be extended to microwave networks of superconducting circuits as well as to photonic networks. In addition, 2D quantum networks, for which the implementation of surface codes becomes possible[78], can also be pursued with the use of four waveguide modes and four corresponding resonator modes.

## APPENDIX: MECHANICAL MODES IN DIAMOND PHONONIC STRUCTURES

### I. Calculations of mechanical compression modes

For wavelengths much larger than the atomic spacing, mechanical modes in an elastic material can be treated as a continuum field with time-dependent displacement at a point **r**, given by **Q**(**r**, *t*). The field displacement obeys a wave equation

$$\rho \partial_t^2 \mathbf{Q} = (\lambda + \mu) \nabla (\nabla \cdot \mathbf{Q}) + \mu \nabla^2 \mathbf{Q}. \tag{7}$$

where $\rho$ is the density of the material. The Lamé constants

$$\lambda = \frac{\nu E}{(1+\nu)(1-2\nu)}, \quad \mu = \frac{E}{2(1+\nu)} \tag{8}$$

are expressed in terms of the Young's modulus $E$ and Poisson ratio $\nu$.

We determine the frequencies and field patterns of the normal modes by solving the corresponding eigenvalue equations using finite element numerical calculations. The material properties of diamond used are $E$ = 1050 GPa, $\nu$=0.2, and $\rho$ = 3539 kg/m$^3$. All structures under study have mirror symmetries, as illustrated in Fig. A1. The solutions of the wave equations will thus be eigenmodes of the symmetry operations. We organize the solutions as even or odd under reflection $R_j$ about a plane perpendicular to the coordinate axis $j = x, y, z$. The specific symmetries of the structure are $R_y$ and $R_z$. All modes considered in this work have even symmetry under $R_z$ (these modes are referred to as symmetric modes). Figure A1 shows the displacement patterns of the third and fourth order compression modes of the thin diamond plate discussed in Fig. 2a of the main text.

## II. Resonator-waveguide Coupling

We describe the coupling between the plate resonators and the phononic crystal waveguides by using a standard coupled-mode theory. The Hamiltonian for a pair of single-mode resonators connected by a waveguide is taken to be

$$H = \sum_n \hbar \left\{ \Delta_n \hat{b}_n^\dagger \hat{b}_n + g_n \left[ \left( \hat{a}_1 + (-1)^n \hat{a}_2 \right) \hat{b}_n^\dagger + h.c. \right] \right\}, \tag{9}$$

written in a frame rotating at the resonator frequency, where $\hat{a}_1$ and $\hat{a}_2$ describe the two resonator modes with the same frequency, $\hat{b}_n$ describes the waveguide modes, $g_n$ is the resonator-waveguide coupling rate, and $\Delta_n$ is the frequency difference between the waveguide and the resonator modes. The sign difference on alternating modes reflects alternating symmetry of the eigenmodes in the waveguide. For a waveguide of length $L = 120$ $\mu$m, numerical simulations of the diamond waveguide structure used in this study give a mode spacing about 30 MHz. In the limit that $g$ is much less than the mode spacing, only the resonant or nearly resonant waveguide mode $\hat{b}_0$ needs to be considered.

In the limit of a single waveguide mode, the (unnormalized) eigenmodes are $\psi_0 = a_1 - a_2$ and $\psi_\pm = 4gb_0 + (\Delta_0 \pm \Lambda)(a_1 + a_2)$, with corresponding eigenvalues $\lambda_0 = 0$, $\lambda_\pm = \frac{1}{2}(\Delta_0 \pm \Lambda)$, where $\Lambda = \sqrt{\Delta_0^2 + 8g^2}$. To determine the relevant resonator-waveguide coupling rates for the phononic network structure, we first calculate numerically the relevant eigenmodes of the full structure. As shown in Fig. A2, the eigenmodes occur as triplets, which arise from the coupling between the unperturbed resonator and waveguide modes. From the frequencies of the given triplet, we can then determine both $g$ and $\Delta_0$, with

$$\Delta_0 = \lambda_+ + \lambda_- - 2\lambda_0, \tag{10}$$

and

$$g = \sqrt{\frac{(\lambda_+ - \lambda_-)^2 - \Delta_0^2}{8}}. \tag{11}$$

For the dimensions of the phononic network used in the main text, the third order compression mode features $g = 9.0$ MHz and $\Delta_0 = -3.4$ MHz, while the fifth order compression mode features $g = 3.1$ MHz and $\Delta_0 = -1.9$ MHz. Further fine tuning of the resonator dimensions can reduce $\Delta_0$ to be much smaller than $g$. The coupling rate can also be tuned or tailored by shaping the contact area between the plate resonator and the phononic crystal waveguide.

In the single-waveguide-mode limit, the eigenmode $\psi_0$ should have no contribution from the waveguide mode. As can been seen from the displacement patterns shown in Fig. A2, there are still discernable contributions from the waveguide, which arise from the coupling of the resonators to the adjacent waveguide modes such as $b_{\pm 1}$. In order to avoid the coupling to multiple waveguide modes, the waveguide mode spacing needs to far exceed the waveguide-resonator coupling rate. For diamond-based phononic network, relatively short waveguides are preferred due to the relatively small size of the diamond membranes available. Note that in the limit of long waveguides (i.e. with $g$ much greater than the mode spacing), quantum state transfer schemes similar to those proposed for optical networks can be used[21,22].

## ACKNOWLEDGEMENTS

This work is supported by AFOSR and by NSF under grants No. 1606227 and No. 1641084.

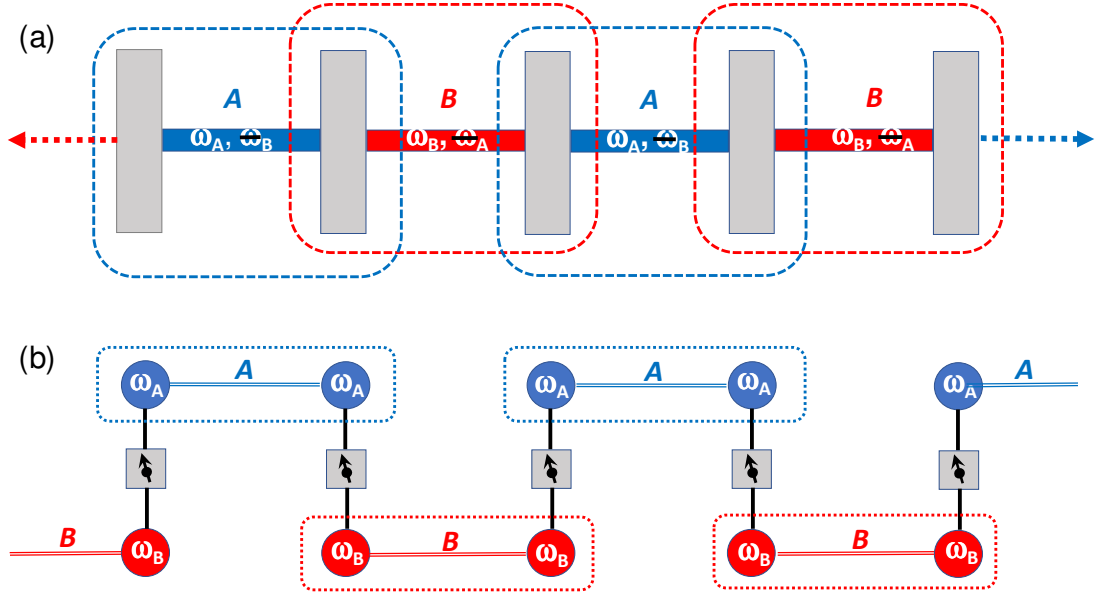

**Figure 1** (a) Schematic of a quantum network with alternating waveguides. Each mechanical resonator (the rectangular plate) couples to two different phononic waveguides, $A$ and $B$. Propagation near frequencies $\omega_A$ and $\omega_B$ is allowed and that near frequencies $\omega_B$ and $\omega_A$ is forbidden for waveguides $A$ and $B$, respectively. (b) In each resonator, a spin system couples selectively to two resonator modes with frequency $\omega_A$ and $\omega_B$. As indicated by the dashed-line boxes, any two neighboring resonators and the waveguide between them can form a closed mechanical subsystem. The network thus consists of closed mechanical subsystems coupled together via the spins.

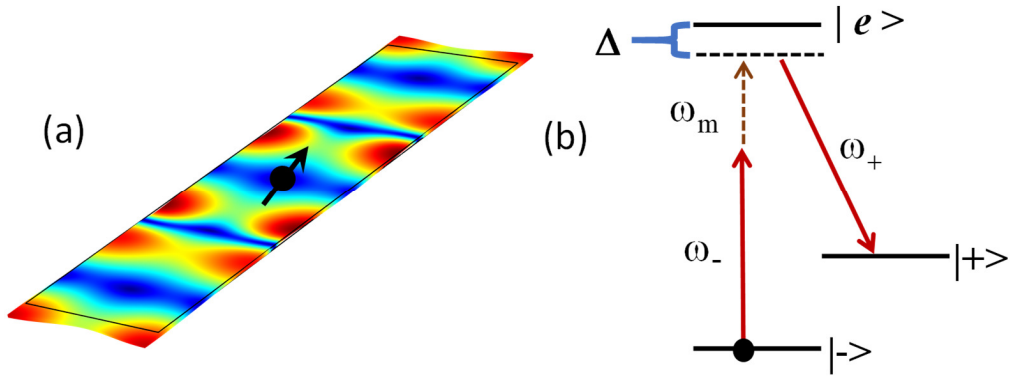

**Figure 2** (a) Displacement pattern of a fifth order compression mode in a thin rectangular diamond plate with dimension (27, 8, 0.3) μm. (b) Schematic of a spin qubit coupling to a mechanical mode with frequency $\omega_m$ through a resonant Raman process, driven by two external optical fields with frequency $\omega_+$ and $\omega_-$. We can couple the spin qubit to a given mechanical mode by choosing a suitable detuning between $\omega_+$ and $\omega_-$.

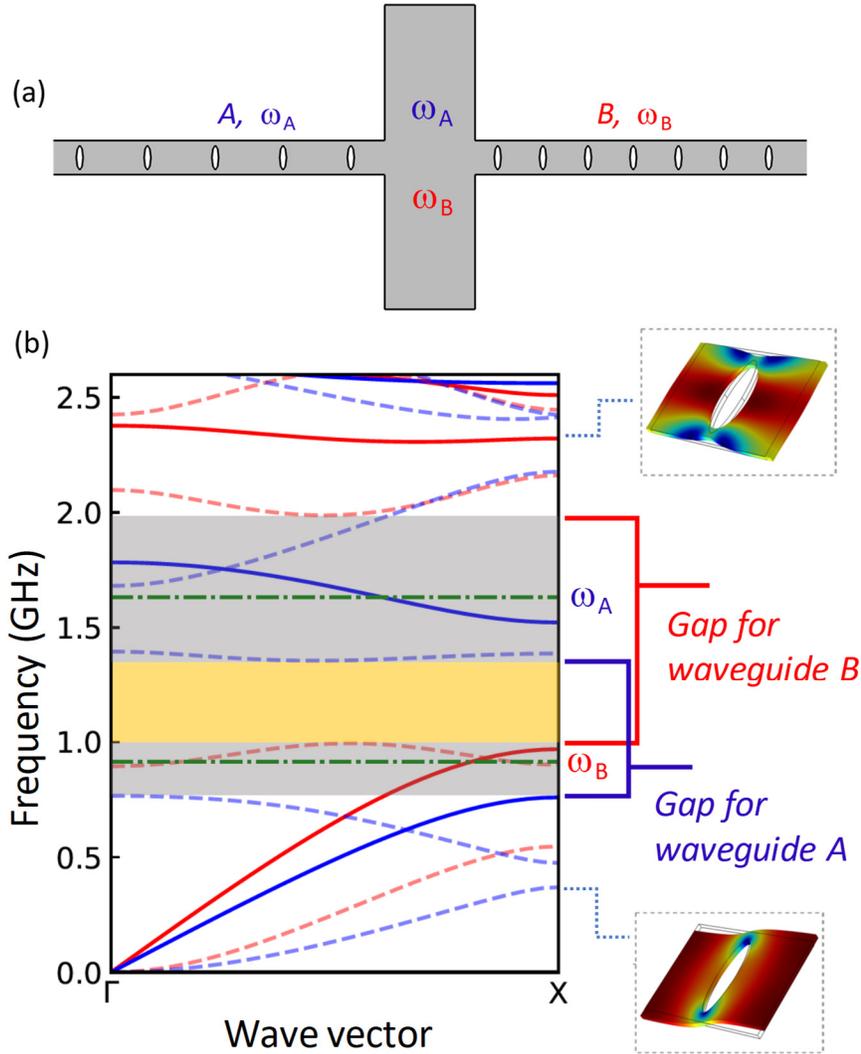

**Figure 3** (a) A mechanical resonator couples to two phononic crystal waveguides with a width of 3 μm and a period of 6 μm (waveguide A) and 4 μm (waveguide B). For the elliptical holes in the waveguides, the minor (major) axes are 0.6 (2.2) μm. (b) Phononic band structures of the two waveguides. Each features a bandgap. Blue (Red) lines: Waveguide $A$ ($B$). The grey (yellow) shaded areas show the non-overlapping (overlapping) regions of the two bandgaps. Solid (dashed) lines are for modes with displacement patterns that are symmetric (antisymmetric) about the plane that bisects and is normal to both the waveguide and the resonators. Dot-dashed lines indicate the frequencies of the two resonator modes, $\omega_A$ and $\omega_B$, used to couple to the respective waveguide modes.

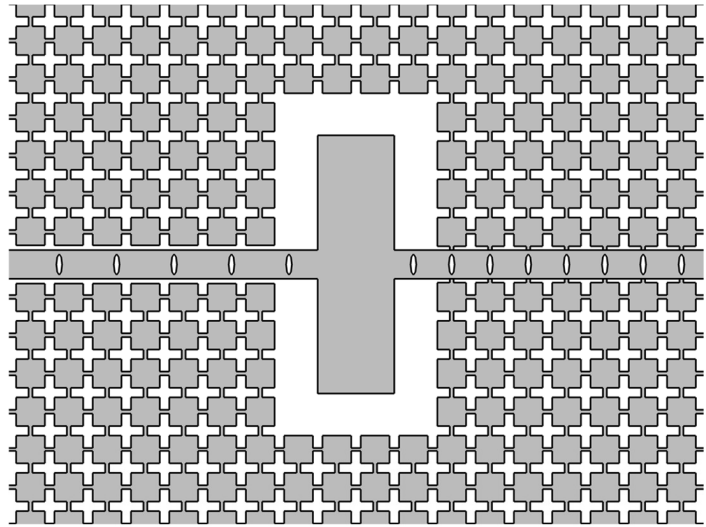

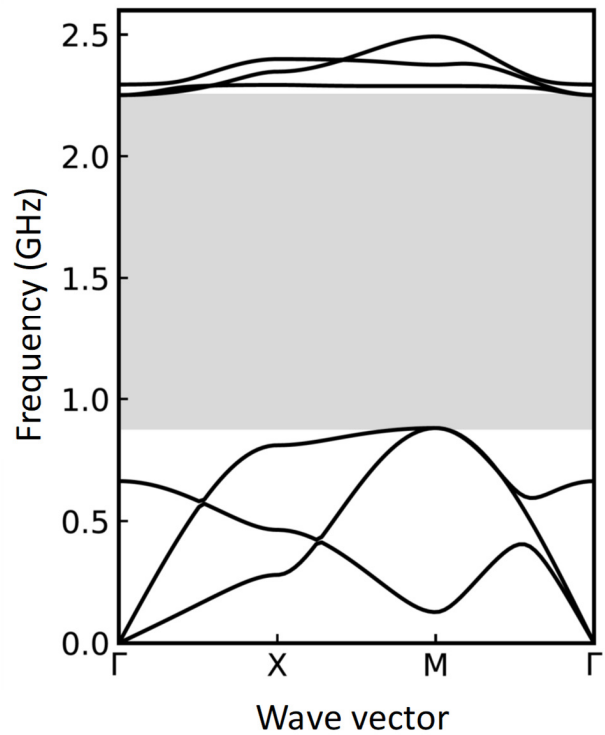

**Figure 4** Top: The building block of a phononic network embedded in a square phononic crystal lattice with a period of 4 μm. The side length of the squares is 3 μm. The connecting bridges have a length of 1 μm and width of 0.4 μm. Bottom: Phononic band structure of the 2D lattice. Only symmetric modes are shown.

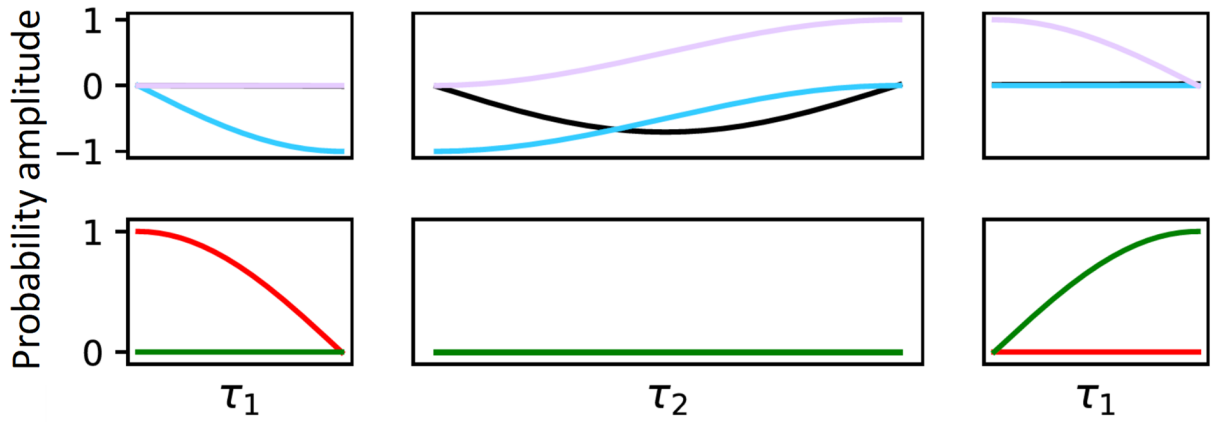

**Figure 5** Time evolution of the mechanical and spin systems with $G/g = 100$ during the three successive swaps of the state transfer between two single spins, with the same peak value $G$ for both $G_1$ and $G_2$. Top: resonator mode 1 (blue), resonator mode 2 (purple), and waveguide mode (black). Bottom: spin 1 (red) and spin 2 (green).

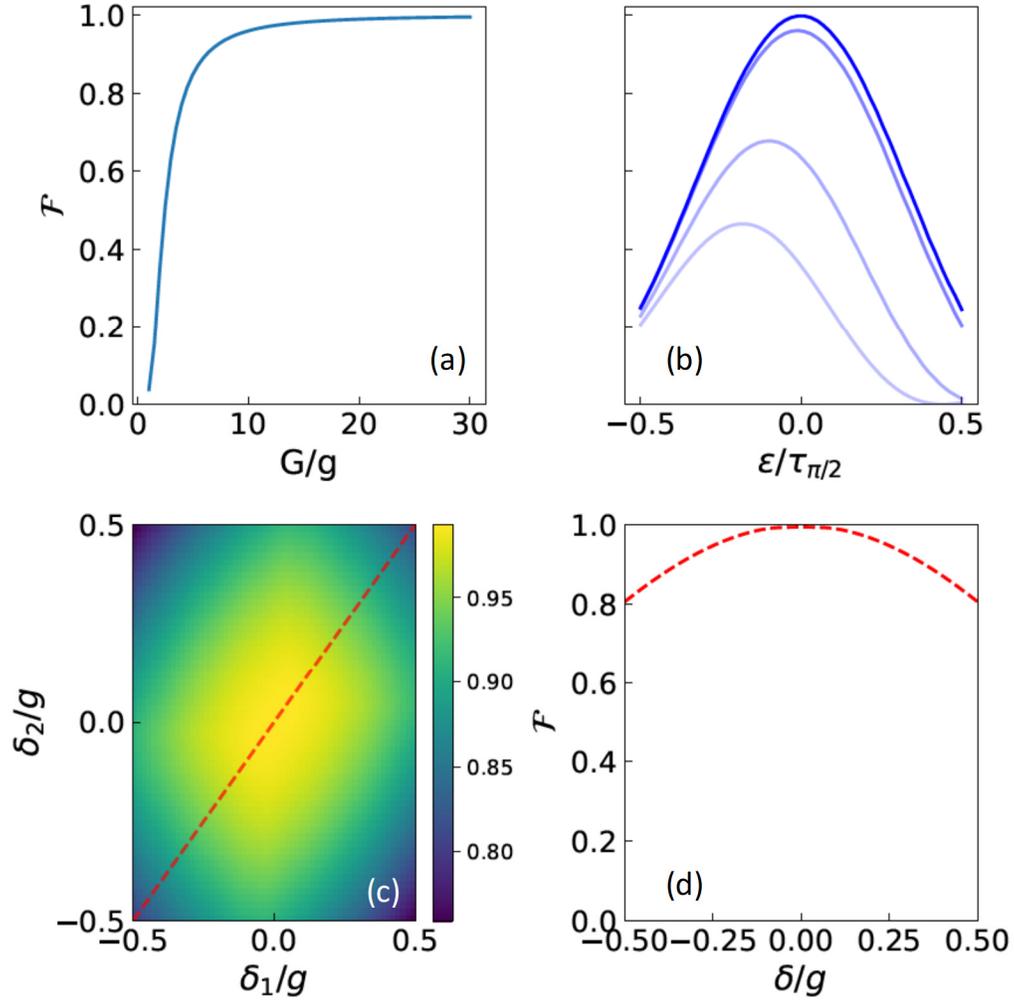

**Figure 6** (a) Lower bound on fidelity for the triple-swap quantum state transfer between two single spins, calculated over all initial spin states and with an initial mechanical state |0, 0, 0⟩, as a function of $G/g$. T=0 K and no decoherence processes are included. (b) As a function of the deviation from the $\pi/2$ pulses. From top to bottom, $G/g$=50, 10, 3, 2. (c) As a function of the detuning, $\delta_1$ and $\delta_2$, between the waveguide and the two resonator modes, with $G/g$=25. (d) As a function of the detuning $\delta_1=\delta_2=\delta$. The result is also marked with the red dashed curve in (c). Ideal pulse duration and detuning are used unless otherwise specified.

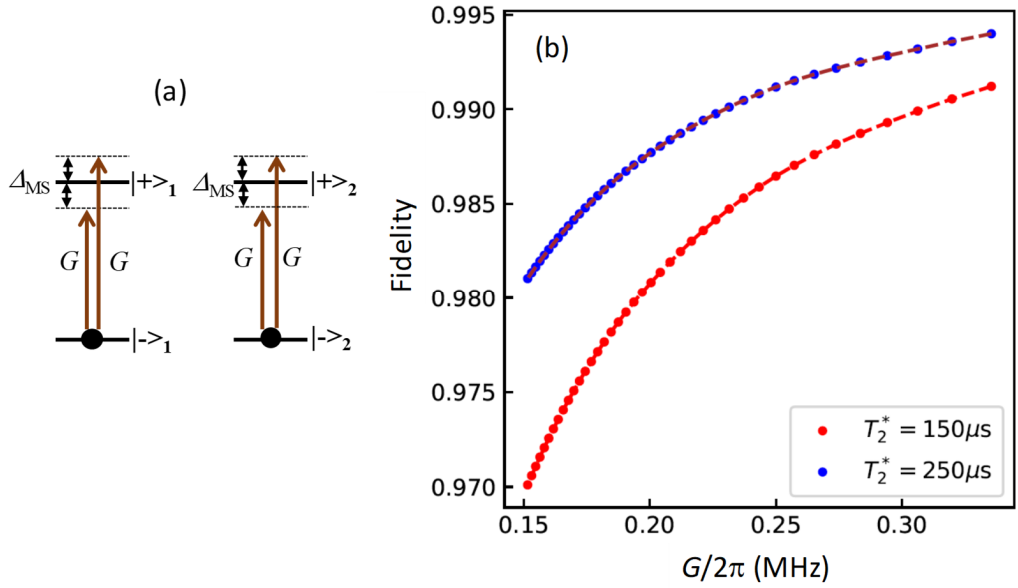

**Figure 7** (a) Schematic of the coupling scheme for the Molmer-Sorensen gate between two single spins initially in state |-->. (b) Fidelity of the Molmer-Sorensen gate at T=0.5 K as a function of $G$, with the mechanical modes initially in a thermal state. Other parameters used are discussed in the text. The dashes lines are a guide to the eye.

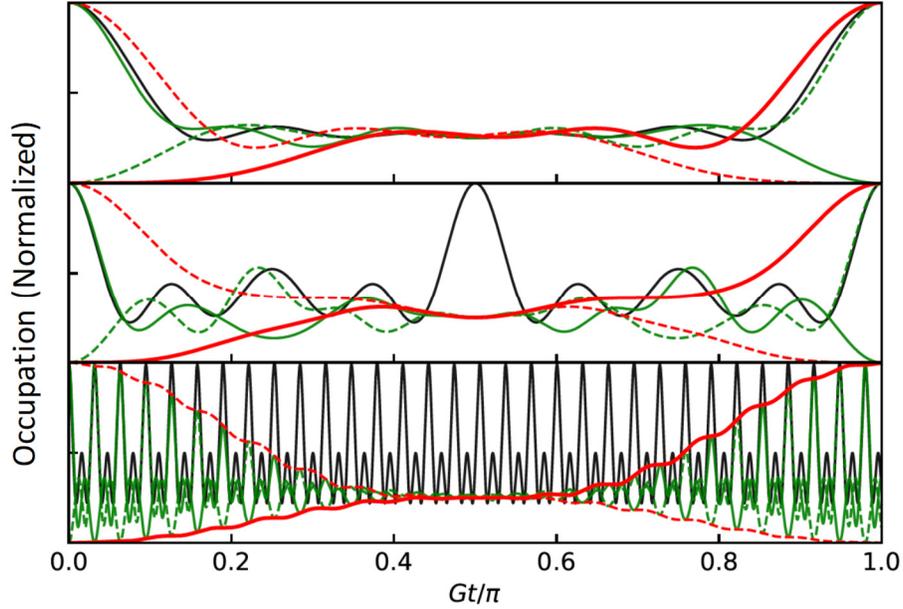

**Figure 8** Time evolution of the mechanical and spin-ensemble systems under a constant spin-mechanical coupling. At $t$=0, the occupation in $\hat{S}_1$, $\hat{a}_1$, and $\hat{b}$ is 1 and that in $\hat{S}_2$ and $\hat{a}_2$ is 0. Red lines: two spin ensembles. Green lines: two resonator modes. Black line: the waveguide mode. Top panel: $\Gamma/G$=2. Middle panel: $\Gamma/G$=4. Bottom panel: $\Gamma/G = \sqrt{1001}$. For both the top and middle panels, the complete state swap between the spin ensembles is accompanied by that between the resonator modes.

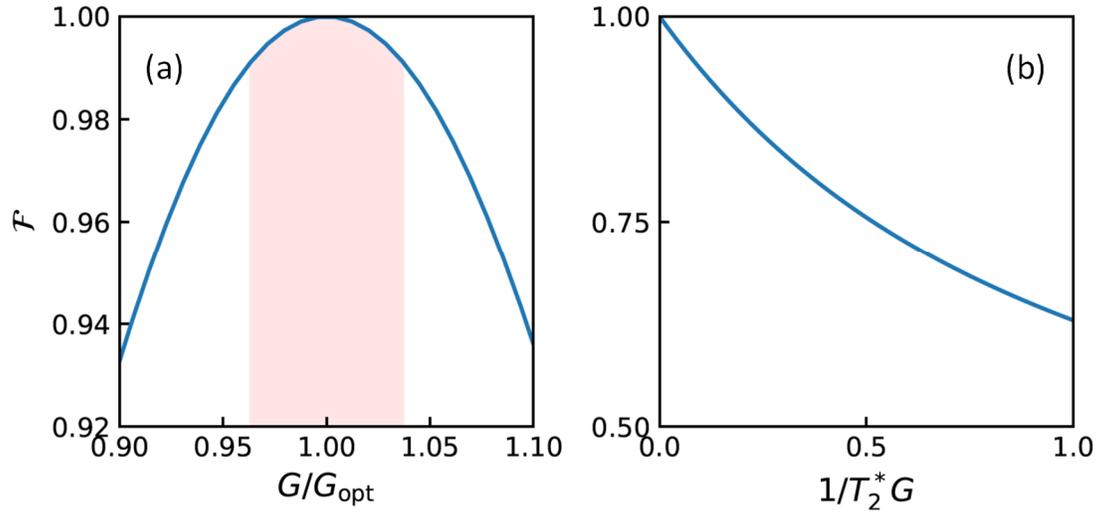

**Figure 9** (a) Lower bound on fidelity for the state transfer between two spin ensembles as a function of $G$, with $\Gamma/G_{opt}=4$ and with the initial occupation in $\hat{a}_1$, $\hat{b}$, and $\hat{a}_2$ given by 1, 1, 0, respectively. T=0 K and no decoherence processes are included. The shaded area indicates the range, for which a fidelity of 0.99 can still be achieved. (b) As a function of the spin dephasing rate, with $G/2\pi =0.1$ MHz, $g/2\pi=9.1$ MHz, and $\delta=0$.

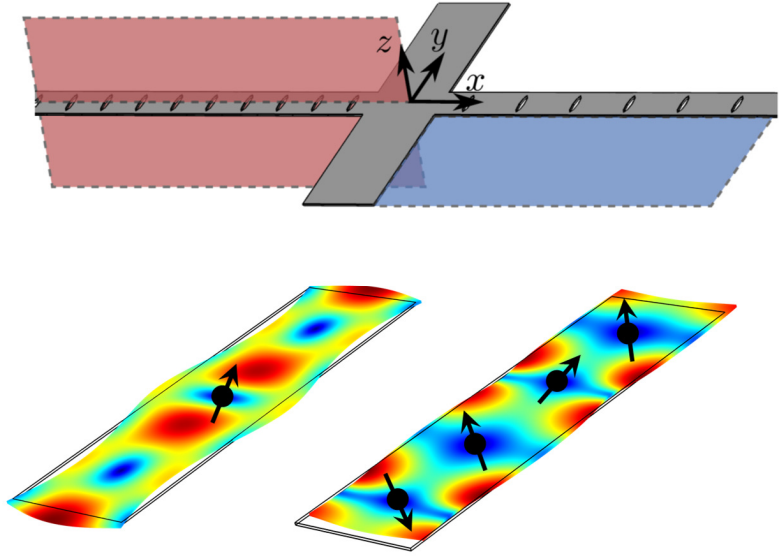

**Figure A1** Top: The reflection symmetry planes of the phononic network structure. The blue and red planes correspond to $R_z$ and $R_y$, respectively. Bottom: Displacement patterns of the third order compression mode (left), with even $R_y$ symmetry, and fourth order compression mode (right), with odd $R_y$ symmetry. The positions of the communication and computation qubits are schematically indicated in the third and fourth order modes, respectively.

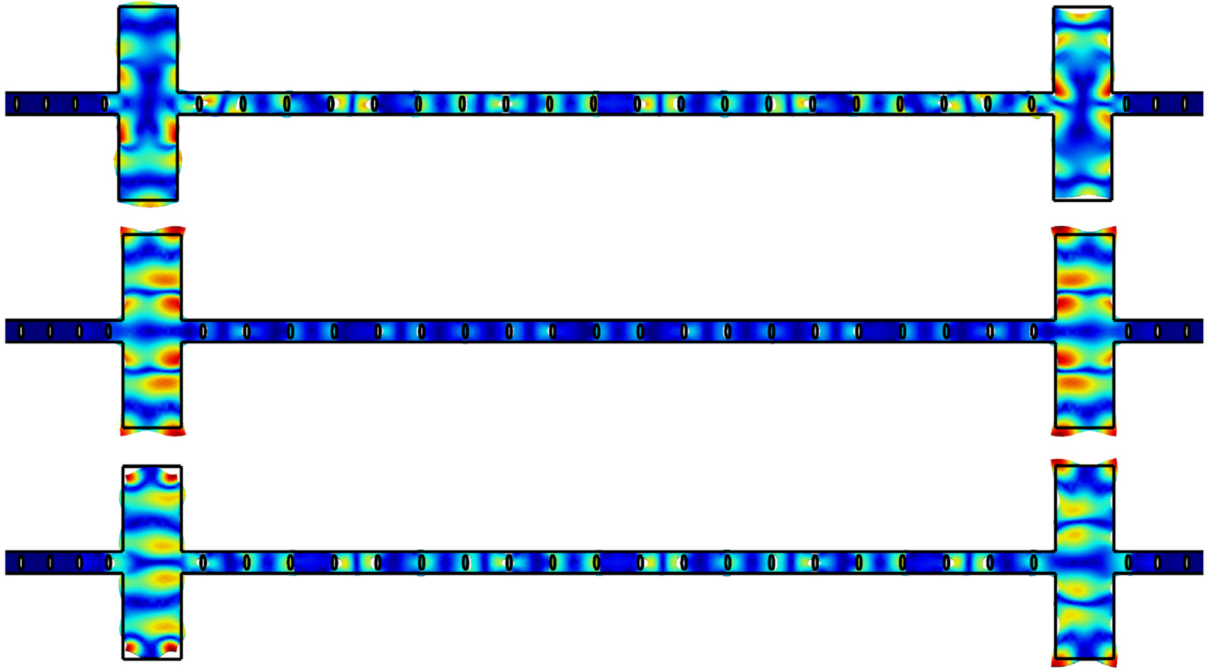

**Figure A2** Displacement patterns of three normal modes in a closed mechanical subsystem. The parameters used are the same as in Fig. 3. The frequencies are (1.6737, 1.6791, 1.6826) GHz from top to bottom. The triplet arises from the coupling between the fifth order compression modes in the two neighboring plate resonators and the nearly resonant waveguide mode. The array of holes in the waveguide has a period of 6 μm.